# Path to Low-Cost Direct Air Capture

Peter Eisenberger, Columbia University and Equitable Climate Innovations Institute

Matthew Realff, Georgia Institute of Technology

## Abstract

It is now accepted that gigatonnes of Carbon Dioxide Removal (CDR) from the atmosphere are needed to avoid the threat of catastrophic climate change. Direct Air Capture (DAC) is a promising scalable CDR with a relatively small environmental footprint. But questions about DAC cost and energy use remain that are delaying the needed DAC policy decisions to create a mobilization effort like was done in the Manhattan Project and to address the Covid crisis. Global Thermostat (GT) has publicly claimed costs of under $50 per tonne for mature GT technology deployed at a climate-relevant scale[1]. Why this low DAC cost is achievable will be addressed by a simplified analysis of generic DAC costs and using that analysis combined with experimental data to evaluate GT's DAC technology. A detailed cost analysis of different approaches to DAC by the National Academy of Sciences (NAS) found an approach to DAC that had a learning cost limit as low as $25 per tonne GT's DAC technology will be shown in Appendix 1 to have the same performance characteristics of the lowest-cost DAC identified in the NAS study. Thus, like solar costs, DAC costs can be reduced by learning by doing, but in the case of DAC, only one order of magnitude in cost reduction is needed[2].Therefore DAC technology can reach its low learning by doing cost limit  at a scale much smaller than necessary to address climate change. From a climate perspective, current DAC embodiments costs and scale have less relevance than their learning curve cost limit.  While GT's technology has demonstrated the crucial performance parameters to achieve a low cost DAC, no inference should be drawn that other approaches cannot achieve low or lower cost, if they can demonstrate the crucial performance parameters. Continued R&D on those performance parameters is needed. In parallel the six orders of magnitude increase in DAC capacity needed in 20-30 years to minimize the risk of catastrophic climate change requires a global mobilization effort starting immediately. Any further delay will significantly increase the risk of exceeding global warming targets and associated climate impacts.

# Introduction

The cost and energy use of Direct Air Capture (DAC) at the scale needed for climate change has very important policy implications. The decisions about its use need to be made now to be able to reach the needed capacity to address the climate threat on time. Because of this, attempts have been made to project its future costs at scale by using learning by doing. Klaus Lackner et al[3] in their paper showed that the required DAC learning cost reductions were two orders of magnitude less than has been achieved by solar. But that paper also raised the important question of whether learning by doing for DAC could achieve the needed cost reductions like solar or fail like the case of nuclear-generated electricity as its critic's claim. A more recent paper published in Joule by Katrin Sievert et al [4]provided a more detailed analysis of DAC learning by doing cost reductions and concluded that the answer to the question raised by Lackner et al[2] was that learning by doing would not reduce the DAC costs below $100 dollars per tonne and would likely cost much more. They asserted, as did the Lackner paper, that their conclusions had important implications for the use of DAC to address the climate threat. They, like many others, used the Climeworks and Carbon Engineering embodiments for their analysis of solid sorbent and liquid sorbents DAC technologies.

The recent paper assumes that all DAC approaches of a given type are on the same learning curve: one learning curve with some learning rate by doing. The paper does not include learning effects created by the R&D phase of a technology. DAC is still in the R&D phase, with capacities of current units in the thousands of tonnes per year level, and hence can benefit from the learning effects of R&D. The R&D phase can shift one to a learning curve that is lower in cost even at a low installed capacity and continues to have a lower cost as one scales the capacity. This is clearly shown in the article[5], in particular Figure 1. In short, there is an R&D contribution to cost reductions, and learning by doing that is capacity dependent. More generally the capacity driven learning by doing approach ignores the question of the learning cost limit that can result in an overestimation of learning rates for those technology embodiments that have a high cost limit and underestimate the impact of R&D on finding embodiments with a low cost learning by doing limit[6].

This article provides data and analysis to conclude that there is a solid sorbent technology that can reach costs under $100 per tonne to capture $CO_2$ from the air and potentially even under $50 per tonne at scale. Thus, providing critical input on the question raised by Lackner et al on whether DAC can achieve the cost reductions by learning by doing that are needed to make it economically viable. We understand that NET $CO_2$ removal is important for climate change for all DAC technologies and can affect the cost. For comparison purposes we focus on the costs of capture. While uncertainties and risks exist for any attempt to predict future costs the threat of climate change requires, we use the best available information for the basis for policy decisions much like was done to create the Manhattan Project and to do so as soon as possible. The emphasis on scale is critical because to address climate change, we will need an industry whose size is a significant fraction of the current global anthropogenic $CO_2$ emissions, i.e., at the gigatonne per year scale[7].

We, and others, have argued previously [8] [9] that the key to DAC is to view it as a dilute separation process where it is necessary to avoid putting work into the feed stream and instead focus the effort on regeneration. This has been borne out by the prevalent DAC designs that bring the air in contact with a spontaneously reacting sorbent in a low-pressure drop contactor, which is then regenerated. Most energy is spent regenerating the sorbent rather than moving the air through the contactor. Below, we demonstrate the major costs of the system and demonstrate, heuristically, that the cost of $CO_2$ capture, not including further purification, compression, and transportation, can be less than $50 /tonne, consistent with the NAS study (Appendix 1). The critical feature of the contactor is how much of the $CO_2$ is removed for the pressure drop that is incurred. An early analysis of this problem was provided in[10] who analyzed that for the potassium hydroxide solution sorbent, the limiting factor of the absorption rate was the liquid film mass transfer coefficient. They provided a very simple expression for removing CO2 that can be used to benchmark different contactor systems (see Equation 1). The key assumption is that the rate of $CO_2$ delivered to the contactor surface from the bulk air channel is matched by the mass transfer into the "wall" of the device and any reaction rate of the $CO_2$ with the sorbent.

The outcome of using this model is that it is possible with solid porous contactor walls to remove more than half the $CO_2$ from the air stream with flow velocities of 3-5 m/sec and that the pressure drop incurred for such removal rates is under 200 pascals in the contactor itself.

In addition to the pressure drop metric, the absolute rate of $CO_2$ removal is critical, which is why high air velocities are necessary. The system is limited by the amount of $CO_2$ delivered per unit time. An order of magnitude lower air velocity might increase the fraction of $CO_2$ removed in an exponentially declining function. Still, it reduces the quantity of $CO_2$ per time in direct proportionality. Maintaining high system productivity is critical to lowering the capital cost and energy used per tonne of $CO_2$ removed, a point we will return to later.

Fans with reasonable air flow rates and efficiencies can provide the 100's pascals order of magnitude of pressure drop required for these removal efficiencies. Assume conservatively a contactor pressure drop of 200 pascals and an additional system pressure drop of 200 pascals for routing the air through the contactor. With an air velocity of 5 m/sec and an 80% efficient fan, the work required is 300 kWh/tonne $CO_2$, which for modest electricity costs of $0.025/kWh leads to a cost per tonne of $CO_2$ of $7.5. At scale in ten years, renewable energy is predicted to cost 1-2 cts per kWh and has already achieved those costs in favorable sites[11]. This demonstrates that the cost of moving the air per tonne of $CO_2$ is not a barrier to the capture for solid adsorbents and that a contactor with low-pressure drop at a high velocity of the input air is critical to achieving low cost. Other electrical energy consumption is associated with pulling a vacuum to remove residual air from the contractor prior to regeneration and during other parts of the cycle, such as cooling by flash evaporation. Still, these contribute a relatively small component to the energy cost, given their brief duration and low flow rates. Overall, the electrical energy cost of the fans and vacuum systems could be around 400 kWh/tonne or $10 /tonne.

The second major component of the system's energy consumption is regeneration, which releases $CO_2$ from the adsorbent. The regeneration for solid adsorbents requires that the system is heated to moderate temperatures in the range of 90-100C and that the heat of desorption of roughly 80 kj/mol $CO_2$ is provided, about 2 gigajoules per tonne of $CO_2$ . The high heat of desorption is an almost inevitable consequence of the requirements to have reasonable removal rates at low concentrations and modest temperature swings. Continued R&D on sorbents with lower heats of reaction is needed that have similar $CO_2$ absorption capacities.

A significant issue is whether water co-adsorbed with $CO_2$ must also be desorbed during this temperature swing. The concentration of water in the atmosphere is 100x higher than that of $CO_2$, in the 10,000ppm range, and most adsorbents that have reasonable capacities for $CO_2$ will also adsorb water due to the chemical similarity between the two molecules. The most straightforward way to avoid significant water desorption is always to maintain a partial pressure of water in the desorption process that is equal or exceeds that of the material at the conditions of adsorption. This can be done by using steam to increase the partial pressure of water. This steam plays multiple roles in the desorption process. It can provide the heat necessary to desorb the $CO_2$ by condensing some fraction. It can be used as a sweep stream to move the $CO_2$ out from the module. It can provide a level of concentration swing by diluting the $CO_2$. If steam is used this way, water will condense, and potentially, the water adsorbed in the contactor material increases during desorption. In either case, the water releases heat to the device at roughly 40 kj/mol water.

The second component of heat consumption is the sensible heat load of the contactor material itself. This must be provided every time the contactor is heated, and therefore, maximizing the amount of $CO_2$ captured per cycle per unit mass of the contactor is important to avoid very high sensible heat loads. We can calculate the heat load for a DAC system based on a few assumptions. First, we show below that the sensible heat load of the monolith contactors used by GT, including the coating and embedded sorbent, is similar to that of the heat of desorption of the $CO_2$ and thus will double the heat required to 160 kj/mol $CO_2$, under 4 gigajoules per tonne or roughly 1kWh/kg of $CO_2$. We will use $0.025 per kWh, a reasonable chemical plant price for moderate steam pressures. This provides a reasonable bound on the cost of steam at 100 degrees, which frequently has no other use in the GT process. This leads to $25/tonne for the cost of steam energy.

Thus, the overall cost of the energy for the system is approximately $50/tonne, even if we assume that the steam and electricity are not co-generated. In the cogeneration case of extracting electrical and heat energy from large facilities that produce for much more electricity and heat needed for DAC one could only pay the plant cost of electricity and get the waste heat at very low cost. Thus, the energy could be as low as $10/tonne. Low temperature heat in the form of waste heat is costly to get rid of and poises environmental risks. For cogenerating with nuclear, concentrated solar or geothermal sources the cost of low temperature steam could be an Opex cost as low as zero.

This demonstrates that the energy cost for DAC can be significantly less than $50 per tonne under a reasonable set of assumptions of which the most important is the quantity of water that is desorbed with the $CO_2$. If 10x the water is desorbed compared to the $CO_2$, then using a typical water heat of adsorption of 40 kJ/mol, the energy required will balloon to 480 kj/mol $CO_2$ without considering the sensible heat of the contactor, tripling the heat to 3kWh/kg.

Thus, the path to low energy-cost DAC is through avoiding high-pressure drop contactors, ensuring that the CO2 per cycle is high to avoid large sensible heat penalties, and avoiding desorbing water from the material. All of this is achieved in systems with direct steam heating of monolith contactors that have loadings of $CO_2$ adsorbent materials. The key remaining component of the cost is the capital expenditure.

The estimation of capital expenditure for technologies at scale from early-stage costs is highly uncertain. One approach is to find analogs of the technology that have similar complexity but are already produced at scale. For example, liquid alkali DAC systems are most closely aligned with large chemical plants, so their capital costs might scale similarly to capital costs for ammonia plants. The capital cost of a large-scale greenfield ammonia plant is $1,300-2000 per tonne capacity of ammonia[12]. With a moderate capital return factor of 0.14, this would lead to an annualized capex of $180 to 280/tonne of production. A similar number is reported for ethylene cracking plants. Ammonia and ethylene are bulk chemicals, but their production involves either catalytic reactors at high pressure or furnaces at high temperature and, therefore, are not perfect analogs for the alkali absorption processes that combine new contactor technology with more traditional pellet reactors and lime kilns. The analogy to chemical plants is not appropriate for the modular approach of solid adsorbent DAC. Instead, what should be considered is manufacturing many smaller contractors, each one of which will have a lower complexity. For example, consider a Class 8 truck. Class 8 trucks are manufactured at the rate of 10,000 per year at sales prices of roughly $120,000, and they weigh about 8,000, which places their cost at $15/kg of truck. Similarly, a diesel train engine, manufactured at lower rates, might cost $ 3 million dollars and weigh about 200,000 kg for a similar cost per kg. Assuming that the solid adsorbent contactor device can capture one tonne $CO_2$ per kg of contactor per year, and that the contactor is 10% of the weight of the device, then this leads to a capital cost of $150/(tonne $CO_2$/year) and an annualized capital cost of $21/tonne $CO_2$.

The above arguments are based on a heuristic calculation of costs of energy and capital, and these are not the only inputs to the cost of $CO_2$ capture, where labor and other indirect costs will play a role. However, the order of magnitude of the energy and capital costs is $100/tonne with moderate assumptions on the price of natural gas and electricity and with the potential to be as low as $50 per tonne. What will drive the cost up or down will be primarily the productivity of the system, captured kg $CO_2$ per kg of sorbent, which controls capital costs and indirectly the labor and other operating costs, and the heat input, again a function of the sorbent productivity and the parasitic load of heating the device and associated desorption of water.

In writing this article, we do not intend to disparage the tremendous efforts of other DAC pioneers or the many more recent startups[13]. DAC efforts to date have played a valuable role in contributing to the increased acceptance of DAC as a critically needed technology for climate change. They have provided the basis for recent significant increases in public support[14]. Soon, the economic demand for DAC will be so substantial that a DAC industry will need to address it. The global price of $CO_2$ will vary greatly, allowing many DAC companies to be profitable. However, this paper will focus on DAC's role in addressing the threat of climate change. This distinction between the climate change focus and establishing commercial industry is very important. In addressing a risk to global security where time is critical, one should take technical risks that the current technology readiness level (TRL) process tries to minimize. We argue that these risks are minimal given what is already known about DAC.

The National Academy of Sciences (NAS) studied the costs of various approaches to DAC without connecting the approaches to any specific effort or company[15]. It showed that DAC costs varied greatly depending upon which method was used, varying between $18 -$1000 per tonne in the first version and $25 as the lower limit in the published version. GT's technology is analyzed below using a simplified cost analysis. It will be shown in Appendix 1 to have the same characteristics as the lowest-cost approach identified by the NAS study.

Uncertainty about the costs of DAC and some confusion about energy use are delaying the mobilization effort needed to address the climate change threat. Such delay is ill-advised if we are to have a chance to avoid the catastrophic consequences of climate change. Thus, the objective here is to show that an economically viable cost path to DAC exists and is ready to scale. In doing so, it will help catalyze future capacity increases and cost reductions to reach under $50 per tonne faster to help minimize the risk of catastrophic climate change. Even with that, it will still take a global mobilization effort similar to but greater than that of World War II[16] [17].

## DAC Cost Analysis Introduction

DAC differs significantly from flue gas capture in the initial capture step because it is about 250 times less $CO_2$ concentration. One must move about 3000 tonnes of air for each tonne of $CO_2$ collected. This fact has led many flue gas carbon capture experts to falsely claim that DAC is inherently high-cost. Their reasoning is supported by an analysis of separation technology costs performed by Sherwood (see figure below). The costs roughly scale as 1/concentration because the first step separating a dilute mixture is usually the costliest because handling the dilute feed is expensive, e.g. digging it out of the ground in mining. The first step in DAC is exothermic and air is readily available which makes the first step less expensive than might be predicted by the Sherwood plot. Other actions, like regenerating the $CO_2$ from a sorbent or reactive material, are first order independent of $CO_2$ concentration[8]. Thus, $50 per tonne of DAC $CO_2$ does not follow the Sherwood correlation and differs by orders of magnitude.

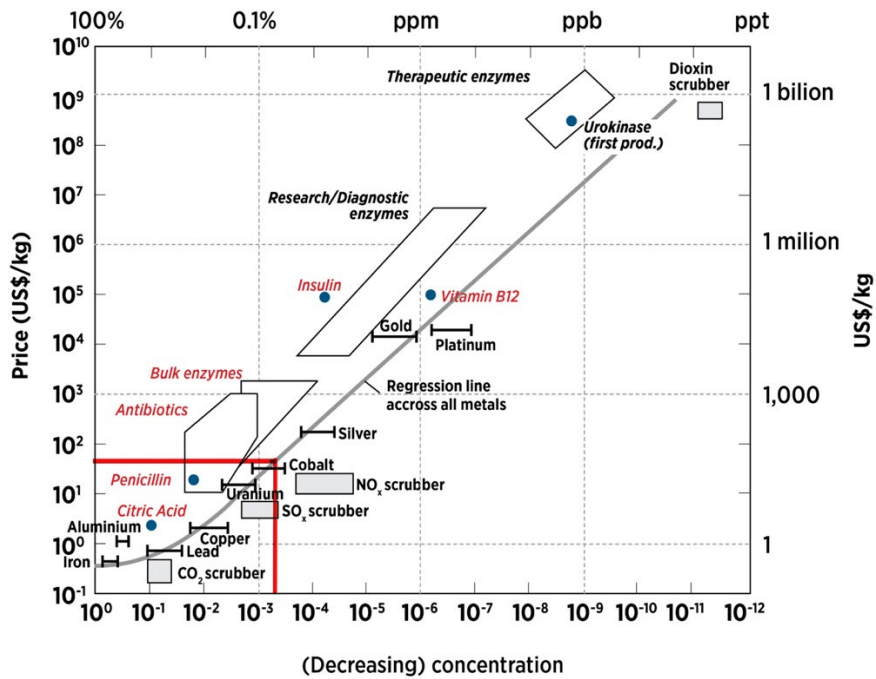

**Figure 1.** Price of selected materials (US$ per kg) as a function of the decreasing concentration of the final product in the initial raw material. The red line shows where the price of DAC $CO_2$ lies if it obeyed Sherwood (thousands of $/tonne) compared to the flue gas scrubber price. <span style="float:right;">Source: Adapted from Dwyer (1984-957) CXSST (1987:22)</span>

This misunderstanding has delayed the recognition of DAC's potential role in addressing the threat of climate change by as much as ten years[18]. The unfortunate situation is that many experts committed to avoiding climate change still advocate against depending on DAC. The reasoning is that because of high costs and energy use, DAC cannot reach the necessary scale to impact $CO_2$ concentrations in any meaningful way. These arguments are delaying the mobilization effort necessary for DAC and climate making it more challenging to achieve the scale needed in the required time[19].

To summarize, the critical performance parameters that define the path to the lowest Annualized CAPEX of DAC are:

1. Maximize the rate in time of $CO_2$ contacting the capturing surface
2. Have kinetics fast enough to capture all the $CO_2$ that contacts the surface
3. Regenerate the $CO_2$ from the capturing material much faster than absorbing it

The second component that characterizes the path for low-cost DAC is low-cost Operating Expense (OPEX). The OPEX is essentially concentration-independent, because the energy is applied to the $CO_2$ in an already concentrated form, and therefore critical performances that define the path to low OPEX are:

1. Minimize the energy used to regenerate the captured $CO_2$
2. Minimize the cost of that energy
3. Minimize the resistance to airflow

## Simplified DAC Cost Model

A simplified cost analysis is used here because the more complex Techno-Economic Analysis (TEA) contains underlying assumptions like depreciation rate, constraints, and risk assessment that make comparisons difficult unless the same approach is used for competing technologies. More fundamental than that, most DAC companies have not provided technical details of their approach because of IP concerns. Even the few that have correctly claimed significant cost reduction in the future do not show how they will achieve it or whether their future costs are above or below the learning curve limit for their path. In short, they fail to answer the question raised by Lackner et al. A simplified approach and the National Academy Study (NAS) analysis of different generic DAC will show DAC under $50 per tonne is possible. It will be shown here with experimental results that GT's approach to DAC has demonstrated the critical performance parameters and in spite of having only laboratory and pilot plant date is on the path to the lowest cost approach that the NAS identified. It is consistent with the importance of R&D for defining a low cost path before scaling to achieve the low cost. As was said above, the distinction between R&D learning and learning by doing is critical, and the situation we find ourselves requires we do both in parallel.

There are, of course, many important technical issues that need to be addressed to have working technology, but if there are no showstoppers, they can be addressed by R&D during learning by doing, as will be discussed below. GT has not identified any showstoppers, nor have other equally mature DAC technologies. Thus, another reason for using a simplified analysis is to focus on the critical DAC performance parameters that enable the lowest learning curve limit identified by the NAS study to be reached.

A high-level heuristic model for the cost for DAC will help identify the critical performance parameters for achieving low cost DAC and the experiments to test them. The variables and parameters are defined below:

| Variable or Parameter | Units | Definition |
|---|---|---|
| $Annual\ Capital$ | $\dfrac{\$/year}{tonne/year}$ | Annualized capital cost per (tonne/year) capacity of the Capture Device |
| $C_A$ | $/m² | Capture Device Capital Cost per m² frontal area |
| $C_E$ | $/m² | Processing Equipment Cost per m² frontal area |
| $C_U$ | $/m² | Contacting Unit Capital Cost per m² frontal area |
| $C_{BP}$ | $/m² | Balance of Plant Costs per m² frontal area |
| $CRF$ | Year⁻¹ | Capital Recovery Factor |
| $E_{AIR}$ | GJ/tonne | Energy to move the air |
| $E_{COLLECT}$ | GJ/tonne | Energy to collect $CO_2$ from contactor |
| $E_P$ | GJ/tonne | Energy for the Process Equipment |
| $F$ | - | Capture fraction of $CO_2$ incident on the frontal area of the device. |
| $O_A$ | $/tonne | Operating cost per tonne of $CO_2$ captured |
| $O_E$ | $/tonne | Energy cost per tonne of $CO_2$ captured |
| $O_L$ | $/tonne | Labor cost per tonne of $CO_2$ captured |
| $O_M$ | $/tonne | Maintenance cost per tonne of $CO_2$ captured |
| $O_S$ | $/tonne | Sorbent cost per tonne of $CO_2$ captured |
| $v$ | m/sec | Velocity of air |
| $\rho_{CO_2}$ | Tonne $CO_2$/m³ air | Density of $CO_2$ in incoming air |
| $T_{CO_2}$ | Tonne/m²/year | Tonnes of $CO_2$ captured per m² frontal area per year |
| $Y$ | Seconds | Seconds per year the Capture Device operates |

The *Annual Capital* for DAC is primarily determined by the number of tonnes of $CO_2$ captured per m² of frontal area of the capture device, the capital cost per m² of frontal area, and the capital recovery factor $CRF$ per year.

$$Annual\ Capital = \frac{1}{T_{CO_2}} \times C_A \times CRF$$

$$T_{CO_2} = (F \times \rho_{CO_2} \times v)Y$$

The tonnes per year per m² of contactor are calculated from the density of $CO_2$ in the air, the capture fraction, the velocity of the air incident on the front of the contactor and seconds per year that the contactor will operate. The units are expressed as ($/year) per (tonne/year) to indicate that this is the cost of building the capacity to capture 1 tonne per year amortized over the lifetime of the contactor.

$C_A$ includes the cost of the contacting unit $C_U$, processing equipment $C_E$, and balance of plant costs $C_{BP}$, all divided by the frontal area of the contactors which is used as a normalization.

$$C_A = C_U + C_E + C_{BP}$$

$CRF$ varies greatly depending on the financial practices of whom is choosing it and so will not be a focus of this simplified analysis.

Operating Cost per Tonne, $O_A$, is to first to order determined by the cost of energy used per tonne(T) of $CO_2$ collected, $O_E$, the maintenance costs per tonne collected $O_M$, other operating costs such as labor, $O_L$, and finally the cost of replacing the sorbent, $O_S$, which could degrade during operation depending on the sorbent type.

$$O_A = O_E + O_M + O_L + O_S$$

The energy use has three main components, the energy to move the air during capture, $E_{AIR}$, the energy for collecting the $CO_2$, $E_{COLLECT}$, and the energy for running the process equipment, $E_P$. The cost per unit of energy can be different depending on the type of energy used for each of them. Typically, the energy to move the air is provided by fans and hence is electrical. The collection energy could be heat provided by steam, hot water, direct solar or electrical joule heating or direct electrochemical reduction of a species as well as electrical energy to run any vacuum pumps used to reduce pressure or remove air from the contactor prior to the collection phase. In a mature technology, $O_M$ is taken to be a small fraction of $C_A$ and, therefore, will benefit from the lower Annualized CAPEX, so it will not be included in this simplified DAC cost analysis. The labor used in operation of the technology cannot be exactly specified for a scaled technology that will be highly automated and part of a larger facility that can help reduce labor costs. But most importantly the purpose of this paper is not to determine the dollars per tonne. By making the target under $50 per tonne, doubling of the theoretical limit identified by the NAS study, one can easily accommodate the cost of operations since it is an even lower amount than maintenance[20].

The generic impact on reducing DAC costs per tonne of $CO_2$ collected of

- A. Low resistance to airflow through the contactor
- B. High surface area per volume for the contacting surface
- C. High rate of $CO_2$ contacting the surface
- D. Fast $CO_2$ kinetics, mass transfer rate in the contactor to find the sorbent that will capture it
- E. Use of low-temperature heat to minimize the cost of energy and reduce the burden on primary energy use
- F. Maximize the time of adsorption

will be analyzed below and shown to be the critical performance parameters defining the lowest cost path in the NAS study. GT technology will be shown to satisfy A – F and thus is on the lowest cost path of under $25 per tonne for DAC identified in the National Academy of Sciences report on DAC presented in Appendix 1. In all stages of DAC technology, fast kinetics is crucial to low-cost DAC because $CO_2$ is very dilute in the air. Thus maximizing the rate of capture per square meter of contactor frontal area will to first order reduce both capex and opex costs with to second order finding the lowest total cost by tradeoffs between opex and capex.

## Analysis of GT DAC Technology

### Capturing the $CO_2$ in the Air

The GT parallel channel contactor embodiment has the air moving parallel to the surface in the lowest resistance to flow laminar regime (Figure 2). Its contactor has the same parallel channel geometry as the automotive catalytic converter, which must process large exhaust volumes with little resistance. Its properties need, however, to be adjusted to be optimized for the DAC process.

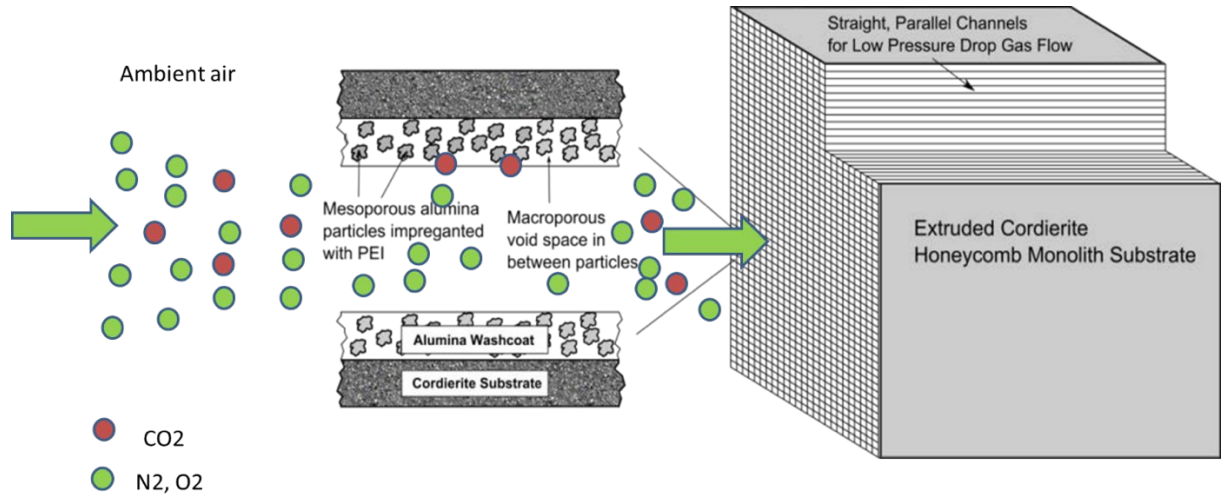

**Figure 2.** GTs contractor. The walls are porous and contain the selective sorbent

The basic properties of such a contactor can be specified by three geometrical parameters

- Size of the channel **S** opening (m)
- The thickness of the walls **W** (m)
- The surface area per volume $S_{av}$ **(/m)** is

$$S_{av} = \frac{4S}{(W+S)^2}$$

where **S** is 1 -2 mm and **W** is a small fraction of **S**, $S_{av}$ is thousands of square meters per cubic meter. Note it is independent of the length **L** of the contactor, but it will be shown below the higher it is, the smaller **L** can be.

The resistance to the air moving parallel to the wall through the channels of the contactor is characterized by the frictional losses at the wall and the resultant pressure drop, $\Delta P$, it experiences. It costs energy to overcome the pressure drop to keep a constant velocity. Air moving parallel to a surface at the velocities used in DAC is in the laminar flow regime, which has a pressure drop given by

$$\Delta P = \frac{32\mu \times v_{channel} \times L}{S^2}$$

Where $\Delta P$ is in pascals, μ is the dynamic viscosity of air (1.84 x 10 -5 kg/(m sec) at 25C) and $v_{channel}$ is the velocity of air in m/sec in the channel which depends on the velocity of the air into the face area multiplied by the face area divided by the face area that is open to flow, equal to $\frac{(D+W)}{D}$.[21]

The use of laminar flow is crucial to avoid unacceptably high pressure drops in the contactor, the friction factor depends inversely on velocity which leads to the above linear relationship with velocity. The pressure drop scales with a velocity raised to a power higher than one for turbulent flow as the friction factor becomes nearly constant instead of scaling inversely with velocity as in laminar flow. Furthermore, the rate of delivery of $CO_2$ to the surfaces of the contactor enables significant removal of the CO₂ without requiring turbulent flow in these narrow channels. The timescale of diffusion of the CO₂ from the center of the channel to the wall is given by:

$$t = \frac{\left(\frac{S}{2}\right)^2}{D}$$

Where D is the diffusion of $CO_2$ in air (m²/sec) and has a value of roughly $2 \times 10^{-5}$, which results in a time of less than one second to reach the wall. Hence, for a contactor of less than 1m in length the required velocity is of order 1 m/sec, which results in laminar conditions in the channel.

The mass transfer rate **K** of the CO₂ perpendicular to the walls for the case of CO₂ in laminar flow air is determined by the correlation of transfer measured with size of the channel and is given by :

**K= 0.059 /S** [m/sec] where **S** is expressed in mm[22] as long as all the CO₂ that strikes the surface is removed.

The fraction, $\theta$, of the CO₂ concentration removed at any time, as the air passes from front to back through the channels, is given by:

$$\theta = KS_{av}$$

Integrating along the length of the channel, the fraction of the incident CO₂ that contacts the walls in a given time τ is given by

$$\eta = \left(1 - e^{-\theta \tau}\right)$$

Where $\tau = \frac{L}{v_{channel}}$ where L is the length of the channel and the above assumptions hold.

For the specific case of the above channel geometry

$$\eta = \left(1 - e^{-K \times \frac{S}{(S+W)^2} \times \frac{L}{v_{channel}}}\right)$$

The tonnes captured per year per m² face area is then given by

$$T_{CO_2} = \rho_{CO_2} \times v \times \eta \times Y$$

One wants to maximize $T_{CO_2}$ with a low-pressure drop. So, the critical performance parameters on the path of the lowest cost DAC **are** low $\Delta P$**,** high $S_{av}$**,** and high mass transfer **K ,** with channel residence times that allow for CO₂ to reach the walls. In addition, the resistance to mass transfer at the wall must match that of the bulk to avoid CO₂ concentrations building up at the wall. This requires both the mass transfer through the porous wall structure and into the material be an order of magnitude faster than K, and that the loading of sorbent does not get close to its equilibrium saturation.

The below measurement data exhibits how the pressure drop $\Delta P$ varies with velocity $v$ for the GT contactor length of 0.15 m. As the analysis above predicted, the $\Delta P\ v$ relationship is relatively linear under GT's conditions, demonstrating the expected laminar flow regime.

**Pressure Drop vs. Velocity**

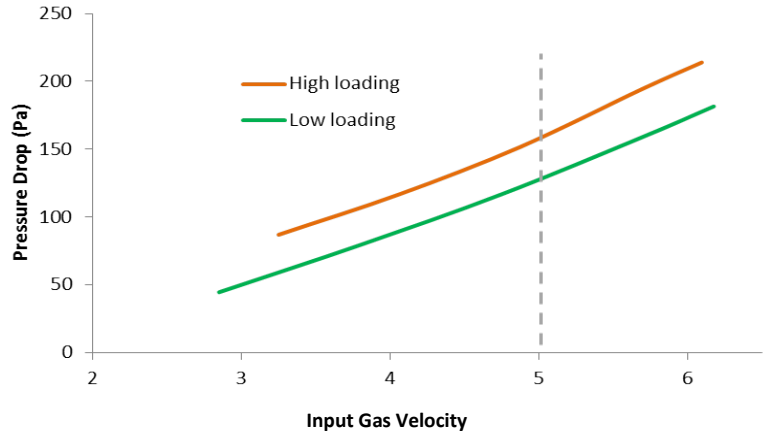

**Figure 3**. – Pressure Drop vs. Velocity measurements for GT monolith contractors at high and low loadings of sorbent material, which results in bigger and smaller wall thicknesses **W** and associated smaller and larger **S** causing a higher or lower pressure drop, respectively as expected (eg (9) . Note at 5 m/s at high loading. The pressure drop is < 150 Pa.

Once the $CO_2$ has contacted the wall , the next physical constraint is that the $CO_2$ must be captured as fast as it impinges on the contactor walls. Otherwise, there will be concentration polarization of $CO_2$ at the contacting surface, reducing the rate of $CO_2$ captured[23]. This reduction in the rate captured over the rate contacting results in lower **η** than predicted by the above model, less $T_{CO_2}$, and, thus, increased $Annual\ Capital$ and $O_A$. This also requires engineering the pore space in the walls to have fast kinetics. The transport to the removal sites is a generic challenge faced by all solid-sorbent-based DAC technologies. As noted earlier, the adsorption capacity per cycle needs to be larger than the amount adsorbed to achieve the fast kinetics needed for low-cost DAC. A similar challenge exists for liquid-based contactors, but $CO_2$ diffusion is as much as 10,000 times slower in liquids than in gasses.

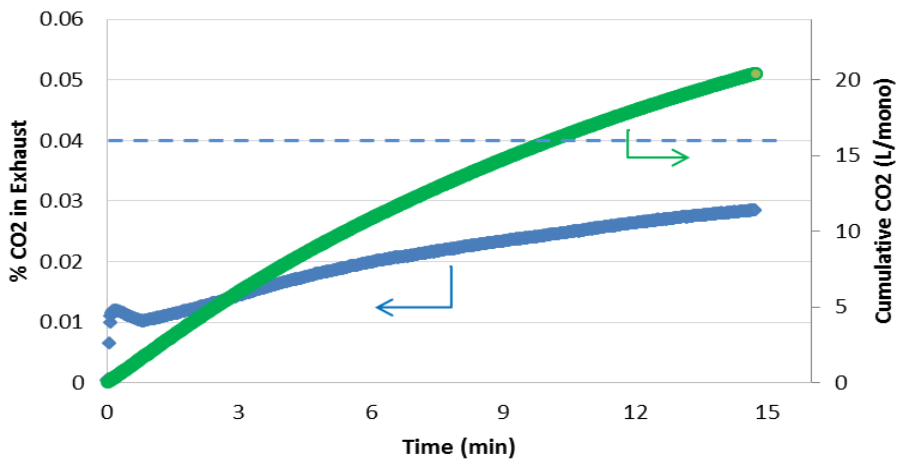

**Figure 4.** – The above data was taken under the conditions ($v$= 5 m/s, **L** = 0.15m, $\Delta P$ < 150 Pa). In 15 minutes, 40L of $CO_2$ at 400 ppm (0.04%) enters the contactor with a frontal area of 36 in² (0.0225 m²). It takes less than .025

seconds at high loading for the CO₂ to pass through the contactor and less than .025 seconds for the CO₂ to diffuse to one of the four walls in the channel.

The above model would predict 24 liters of CO₂ per monolith so the impact of concentration polarization is minimal and can be further reduced. Thus, GT's monolith parallel channel contactor with low $\Delta P$, high $v$, high $\eta$, and small L length performs close to the one used in the NAS lowest cost DAC scenario, maximizing $T_{CO_2}$. It also has a small volume of contactors because of its high $SA_v$ and **K,** thus reducing **CC.** There are other factors that can impact $T_{CO_2}$ because DAC is open to the variations in the air in terms of humidity that can lead to the presence of water that can slow down the diffusion and contaminants that can impact the lifetime of the sorbent. While these can present significant challenges because DAC can be located anywhere, they can be avoided by operating in areas where the humidity optimizes the performance and the contaminants are low. Because these types of issues always emerge in technologies that are done in various environments and which for example the petrochemical industry has successfully addressed via R&D. There is no reason to believe that the same can be done for DAC with an ongoing R&D effort. The failure to include future improvements by R&D will result in predicting higher DAC costs and energy use than can be achieved. Most importantly because time is a real constraint they are the type of risk that one should take and address while scaling.

## Regeneration of Captured CO₂

Generic low-cost DAC technology needs to collect CO₂ as fast as possible because while collecting the CO₂, it is not absorbing CO₂, thus reducing the $T_{CO_2}$. It is an energy-intensive process because it takes energy to remove the CO₂ bound to the sorbent. There are many known mechanisms to desorb the CO₂:

A. Heating to desorb the CO₂ from the sorbent, a temperature swing adsorption cycle, TSA,

B. Creating a vacuum over the sorbent to reduce the absolute pressure, a vacuum swing adsorption cycle, VSA

C. Using gas to effectively sweep the CO₂ off the sorbent by creating a low CO₂ partial pressure, a concentration swing adsorption cycle, CSA,

D. Using membranes by reacting the CO₂ with a mobile sweep phase that is then separately regenerated,

E. Electrochemically

F. Displacing the CO₂ from the adsorbent using a more strongly binding molecule, e.g. water in a humidity swing cycle.

G. A combination of these mechanisms, for example a temperature, vacuum, concentration swing adsorption cycle, TVCSA.

Here, as described earlier, there are two parameters that are the main determinants of energy costs per tonne of CO₂, the collection energy amount **EPT** and the cost of the energy used **CE**. Not all energy costs are the same. The cost of electricity per joule is higher than the cost of heat per joule, and the cost of high temperature heat is larger than that of low temperature (below 100 °C) heat. The second performance objective is the productivity of the module $T_{CO_2}$, which correlates with the capital cost of capture or **Annualized CAPEX**. To maximize productivity the duration of collection and any time spent in-between desorption and adsorption, such as cooling should be as short as possible, since the longer the collection step the less time the contactor spends adsorbing.

GT has demonstrated a combined TVCSA cycle which uses low energy, low temperature -vacuum swing + sweep vapor combination where the low temperature (less than 100 °C to as low as 70 °C) steam heating of the contactor, desorbing the $CO_2$ and sweeping to achieve a low partial gas pressure of $CO_2$ in the contactor during collection whilst maintaining modest temperature increase by pulling a slight vacuum. It has also demonstrated as described below that the process can be carried out without a vacuum step in a continuous version of the DAC process. The rate of heat delivery to and within the walls is very fast because of the latent heat of direct steam condensation on the walls. The thin walls, small **W**, in the contactor make mass transfer by diffusion and heat transfer by thermal conductivity very fast to complement the rapid heat delivery. The time for heat and mass diffusion scale with **W$^2$** and **W**, respectively, where **W** is tenths of millimeters. The second benefit of direct condensation is that any water that is co-adsorbed with the $CO_2$ will not desorb during this step as the water partial pressure will be higher during desorption compared to adsorption.

To return the contactor to a state where it can contact air again requires the module to cool down. The presence of condensed water allows for flash evaporative cooling to occur without exposure of the material to air at high temperature, or in a continuous process by a nitrogen sweep. A second phase of cooling with direct evaporation of water into the air stream happens at the beginning of adsorption. Both of these direct evaporative cooling steps can be performed quickly for the same reasons that direct steaming is fast. Concerns about the interaction of the steam with the sorbent and porous walls is minimized, if not eliminated, by the "coke bottle effect", the $CO_2$ desorbs rapidly from the pores into the channel and provides a counter flow to any water into the channels. The drawback of direct steaming is the potential loss of water from the contactor during the cooling phase by evaporation and the return of the material to equilibrium with the incoming humidity. In certain climate conditions the device with water recovery can actually co-produce $CO_2$ and water.

### GT's Regeneration Process

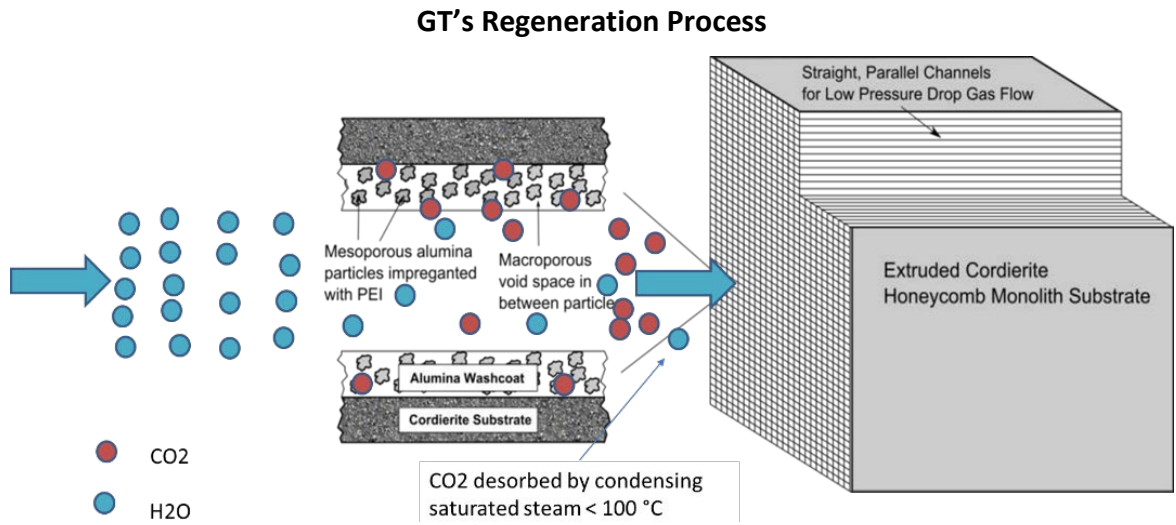

**Figure 5.** GT's $CO_2$ regeneration process

It uses direct contact steam. Both to heat and sweep the $CO_2$ Plug flow, no mixing of upstream and downstream flowing gasses, and can only be achieved in the laminar regime because turbulent flow mixes gasses. Plug flow is vital to achieving sharp steam breakthroughs. It is crucial because if it is not sharp, the steam that exits the monolith without condensing when the steam front approaches the back face of the monolith will be wasted steam and increase **EPT.** Below is data that shows the expected plug flow taken in a batch process regeneration chamber. A plug of oxygen gas introduced into the front face of the monolith has a shape shown at the left of Figure 6. It emerges from the back face of the monolith with the shape shown to the right as measured by an oxygen detector.

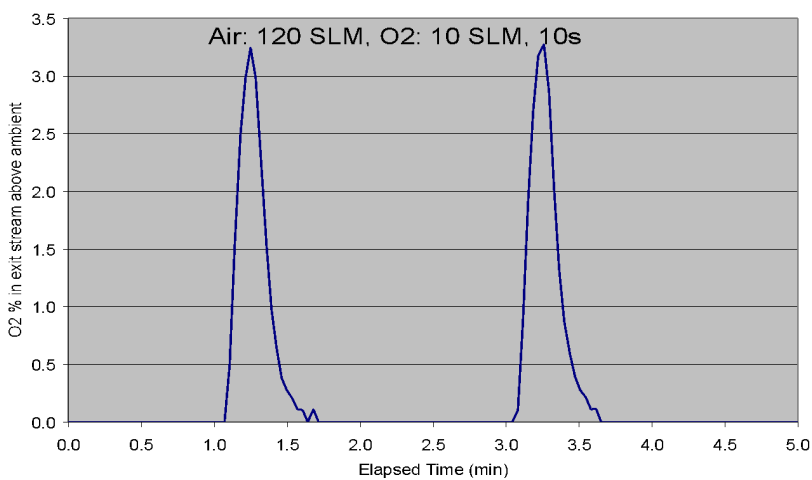

**Figure 6.** Data from a monolith with more than 100 channels

This plug flow under laminar flow follows if one has uniformity of flow over a larger frontal area. The uniformity with which monolith channels can be fabricated means that the flows are well-defined and equal in each channel. The challenge of distributing flow evenly over a large surface area is greatly reduced by using a new version of the continuous GT process. It reduces the area over which plug flow needs to be achieved and makes it more uniform in that area because of the lack of any valves, thus making plug flow easier. The sharpness of the steam front shown in Figure 7 will minimize wasted steam during the collection process by turning off steam injection when the back end reaches temperature.

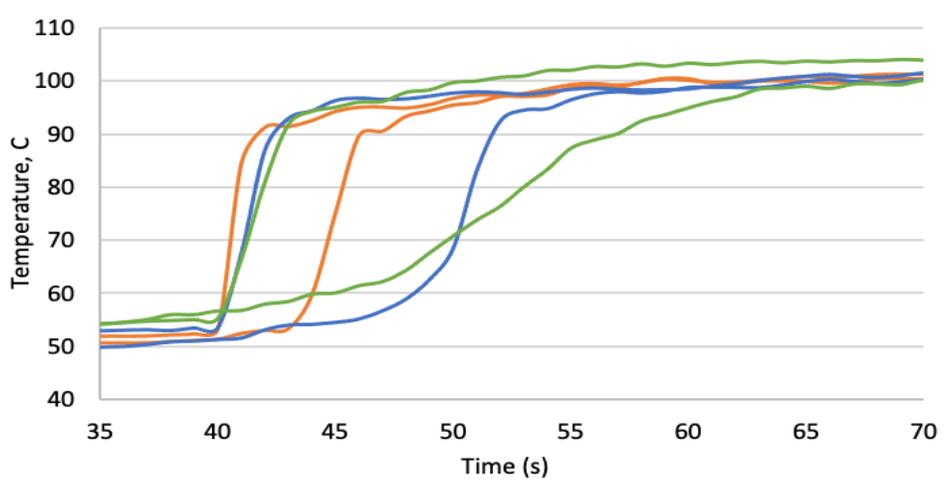

**Figure 7.** Results for $CO_2$ collection during GT's steam sweep collection step. The contactor front temperature rises quickly for all velocities of steam at 40 seconds. The back temperature rises to the right as shown by the second curve of the same color. The delay in steam breakthrough is caused by the steam sweeping the $CO_2$ from the walls which condenses steam inside the channel. The 10cm/sec data is in green, the 20 cm/sec in blue, and the 40 cm/sec in orange.

Another important impact of using direct steam contact is that the steam also acts as a sweep gas, creating a low partial pressure of $CO_2$ in the channels of the contactor. This shifts the Langmuir equilibrium[24] to a lower temperature, increasing the rate at a given temperature at which it leaves the sorbent. Using direct steam contact is also a less energy-intensive way than pumping to create low partial pressures. For example, in adsorption, to avoid slowing down collection, the rate of $CO_2$ moving through the porous walls into the channels and exiting the downstream face of the monolith must be faster than the rate at which $CO_2$ leaves the sorbent. The relatively thin walls and significant macroporosity result in little resistance to flow for the $CO_2$ from the contactor before being swept along the channel by the steam. It also reduces the contactor's thermal mass, reducing the sensible heat needed to raise it to the desorption temperature. The sensible heat of the monolith contactors, their coating, and sorbent for the CO2 capture shown in Figure 4 is determined to be about 3 gigajoules per tonne of $CO_2$ collected. Shutting off the steam when the back end of the channel reaches temperature before steam breakthrough as shown in Figure 7 ensures high steam efficiency. It was shown to be about 90% by measuring the amount of steam energy input compared to the sum of the sensible heat needed to raise the temperature of the loaded monolith and latent heat needed to free the CO2 that was captured. When combined with the performance in adsorption shown in Figure 4 , together with new reduced thermal mass contactors, with heat management while cooling the monoliths after regeneration and with the 2 gigajoules per tonne to free the CO2 from the sorbent means that less than 4 gigajoules per tonne of low-temperature steam is readily achievable. It could be reduced with future R&D to be as low as 3 gigajoules per tonne at scale.  Here again, is a prediction of future performance that can be achieved through R&D while scaling. GT has worked with Corning for close to 10 years. As a result of that interaction Corning has already developed a proprietary contactor with greatly increased porosity and of low thermal mass to make the sensile heat per tonne collected lower than needed to release the $CO_2$ from the sorbent. One can also reduce the $CO_2$ heat of adsorption of the sorbent as other DAC efforts have demonstrated, but with the possible loss of capacity. Heat recovery by vapor compression of the product flow and from water evaporation during cooling can also reduce the net heat required. Again, this is the type of parameter and process evolution that benefits from R&D while learning by doing while scaling.

The performance demonstrated for each stage of the DAC process, together with heat management, has not yet been demonstrated in a single operating unit. The process steps are independent of each other in the sense that each can be optimized without impact on the other stages but of course their impact can be constrained by the other stages. For example, one can optimize the contactor dimensions for pressure drop to surface area but be constrained by the pore structure of the walls to deal with the increased rate of mass transfer of $CO_2$ to the wall by its pore structure. But one can in turn change the pore structure to handle the increased flux of $CO_2$ to the wall keeping the contactor length constant. As will be argued below, from a climate change perspective, because each stage has demonstrated that the approach is on the lowest cost learning path suffices to begin scaling up now. Again, the DAC situation is that it is still in the innovation stage yet the decision to mobilize to scale needs to be made now.  This was certainly the case for the Manhattan and Covid crises.

## Scaling of GT Technologies

GT technology is modular in that one can add additional frontal area and not impact the performance of the previous frontal area and quite literally because the contactor is arranged in standard modular panels. It is also a plug-and-play structure because the mechanical system remains the same as one changes monoliths and sorbents and even the regeneration process by swapping or changing the regeneration box. This is why showing the performance needed in the components makes it likely that they can be combined successfully in a unit. It also can function, as shown above, as a plug flow reactor with all the benefits that it brings: 1. High Conversion per unit Volume 2. Low operating cost) 3. Continuous operation, and 4. Good heat transfer. The monolith contactors are arranged in a monolith panel. The first GT plant at SRI had an elevator to move the monolith panel from adsorption to collecting in ten seconds. This pilot unit also had one collection (regeneration) chamber per contactor module making the capital efficiency of the collecting module with all its piping, valves, and steel only 10%. As mentioned earlier, another benefit of having fast kinetics for DAC is, in this case, a 10 to 1 ratio between capture and collecting times, thus increasing the capital efficiency by 10. GT, in its plant operating at its R&D facility in Colorado, uses a horizontal moving system to have one collection module per 10 monolith panels. It is shown schematically in Figure 8.

### GT's Batch Process Moving System

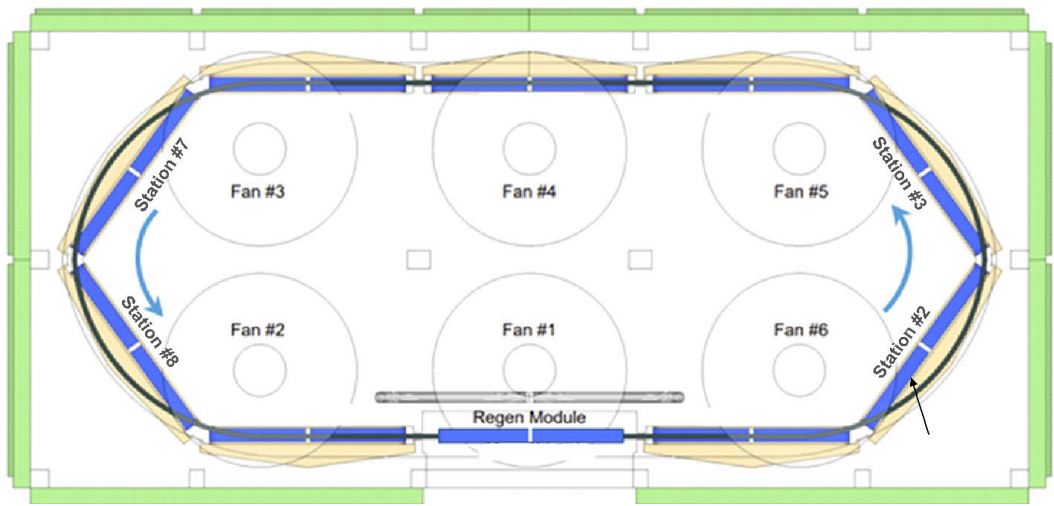

**Figure 8.** GT's Batch Process moving system. There are 10 contactor panels, each moving through 9 stations in adsorption before entering the Regen collecting station

While from the climate perspective, it is ideal to use renewable energy, in the lengthy transition period, natural gas can be utilized. GT has developed DAC$^{+,}$ which captures flue gas $CO_2$ in the same unit that captures $CO_2$ from the air, as shown in figure 9 below.

**DAC+ Block Diagram**

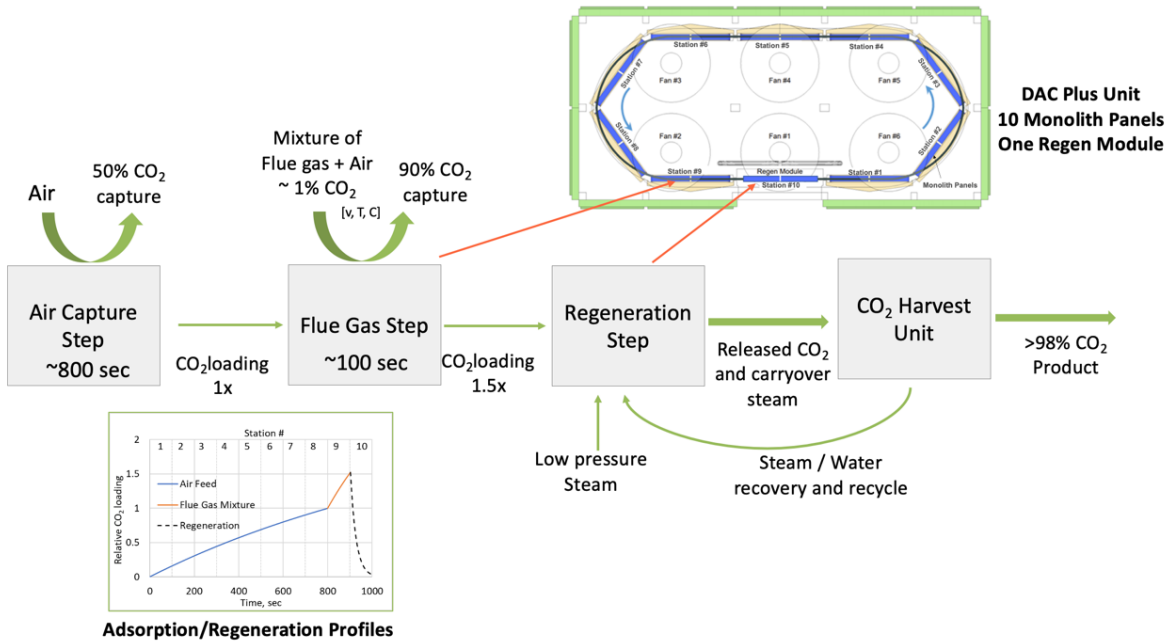

Adsorption/Regeneration Profiles

**Figure 9.** The monolith panels spend 8 stations absorbing $CO_2$ from the air, one station absorbing $CO_2$ from flue gas, and one station collecting $CO_2$. The insert on the lower left represents actual measurements showing a 50% increase in $CO_2$ adsorption in the 9$^{th}$ station.

Natural gas releases 19 gigajoules of energy per tonne of $CO_2$ emitted, and at scale, it takes under 5 gigajoules of energy per tonne to capture $CO_2$ . Thus, one can collect twice as much CO2 from the air than being emitted by the power plant providing the energy for DAC. The increase in $CO_2$ captured is the expected result because of the higher concentration of $CO_2$ in the flue gas. In fact, using GT technology for flue gas is less costly than DAC and notably will be less costly than current flue gas approaches because of its much faster kinetics.

The new GT continuous DAC, CDAC process, enables one to further exploit the fast kinetics in this case by eliminating the need to pump down, stop and start, and open and close valves that batch DAC requires. The CDAC apparatus is shown in figure 10. It also makes achieving plug flow easier, as discussed earlier. It also enables heat management approaches not feasible in a batch process that will reduce the amount of low temperature heat needed. Those processes and many other important details are beyond the scope of this article. But CDAC does enable both the capturing and collecting $CO_2$ stages to benefit from fast kinetics and reduce energy costs. However, it does require some form of sweep gas between the adsorption and desorption phases to remove air from the monolith channels before the steam is introduced and to cool the system down before air is reintroduced. This is because the amine adsorbent is sensitive to high temperature oxidation and cannot see both the partial pressure of $O_2$ in the air and temperatures of regeneration simultaneously. It is on the path of the lowest cost DAC identified by the NAS, as shown in Appendix 1.

## GT's continuous movement system

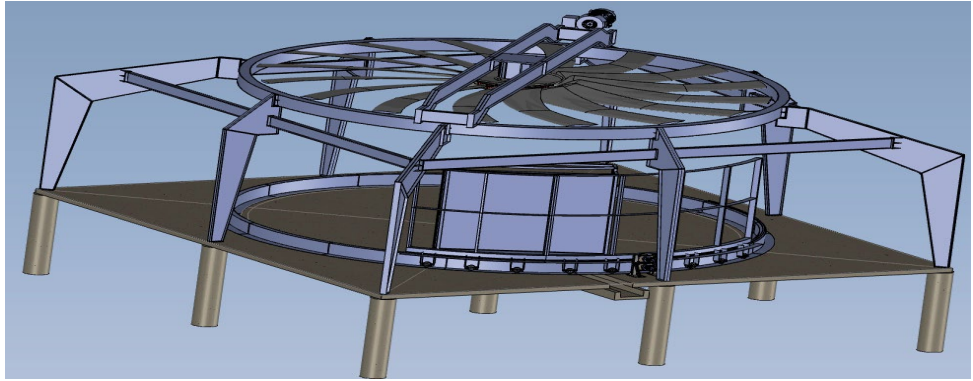

**Figure 10.** GT's continuous movement system - CDAC

How high and wide the monolith panels should be to scale up the GT's CDAC process and how to arrange the scaled-up system so they do not process depleted air is beyond the scope of this paper. But it is something all DAC technologies will need to address to avoid processing already processed air. If the movement system is a closed loop system, the loop can be arranged to have a shape that can adjust to local conditions of wind and other facilities, such as power plants that produce the energy used in the DAC processes and facilities that use the $CO_2$.

Supply chain scaling is critical for any DAC technology because of the large scale needed. To first order rare or costly materials should not be used and the environmental impact should be negligible. We will address water and energy use later. In scaling DAC technology GT has an advantage because its components, monolith, and coatings have already been scaled by suppliers of catalytic converters. Also, those suppliers will have spare manufacturing capability because of the decrease in demand expected from the switch to electric vehicles. Finally the component materials other than the sorbent are all readily available and largely unaffected by use. The sorbent is an animated hydrocarbon that is all used for other commercial purposes and synthesized from readily available materials. In fact, it is components like motors and heat exchangers and fans needed for all DAC processes that will have to scale for a removal market orders of magnitude more than current use. This is a very important reason to start scaling DAC now to give those manufacturers of the components the confidence to start now to scale their production capability. More generally as in World War II supply chains will be most impacted by the mobilization efforts undertaken to meet the climate threat in the time needed[25].

## Learning Curve Cost Reductions

GT DAC and DAC generally are at the beginning of their respective learning curves. They are in the innovation stage where R&D can put you on a lower cost learning by doing curve that depends upon installed capacity. As is now very clear and has been dramatically and recently demonstrated by solar energy, all operating technologies have their costs reduced as one does more of them. But in the case of DAC, only one order of magnitude in cost reduction is needed compared to three orders for solar from its initial conception as shown in Lackner et al [26].

Since a major assertion of this paper is that R&D learning has been and will continue to be critical we will use the standardized approach for the component that depends upon capacity. The paper by Sievert

Katlin et al has done a more detailed analysis (ignoring the R&D component) but their detailed analysis of learning rates for the components that are capacity oriented cost reductions does not reach a different conclusion than the learning rates used below. In fact because one is on a lower cost learning curve even lower learning by doing learning rates reach costs of $100 and $50 dollars per tonne significantly before the gigatonne scale of installed capacity is reached as shown below.

Wrights Law (Wright, T. P. (1936). "Factors Affecting the Cost of Airplanes [27] [28] is appropriate for numbering up – making more units as opposed to increasing the scale of each unit. It will apply to the sorbent apparatus and moving system and components like fans that can be mass produced. The formula for the cost $C_D$ of the modular unit as a function of the number of units N expressed in terms of $N = 2^D$ where D is the number of doublings of capacity with a learning rate LR and a cost reduction per doubling of capacity b where 1-b= LR is given by Wright as,

$$C_D = C_1 \times N^b = C_1 \times 2^{(D \times b)}$$

**14** $C_D = C_1 N^{(*\ln_2(b))} = C_1 2^{(D *\ln_2(b))}$

The learning rate of the chemical industry is .17. This is based on economies achieved by building larger chemical plants, scaling up, rather than by producing more modules, while for the solar industry which scales out, the rate is 0.285.

Using the model for the case, C1 =$200 LR = .25, $50 per tonne will be reached by only 6 doublings of installed capacity. If one chooses C1= $400 and LR=.25, one reaches $50 per tonne in 9-10 doublings of capacity. This shows that for the initial capacities of DAC today of 1000 tonnes per year, that at 1 million tonnes, which is 10 doublings from 1000, that this level of cost could potentially be achieved either by scaling up or by scaling out. The caveat, in both cases, is that the $50 dollars per tonne DAC would be reached only if they are above the learning curve limit of the path that the specific DAC technology is on. In other words, the minimum cost must be below $50 a tonne, which we have established is feasible based on arguments above.

GT's modular technology can capture both the benefits of scaling up and out. As the GT DAC technology is deployed with larger capacity (scaling up), then one reduces the capital cost of $C_E$ **and** $C_{BP}$ costs which follow the cost reductions of scaling up.

The cost $C_E^2$ and $C_{BP}^2$ of a plant with $T_{CO_2}^2$ capacity compared to the cost of a plant $C_E^1$, $C_{BP}^1$ with capacity $T_{CO_2}^1$ is given by

$$C_x^2 = C_x^1 \times \left(\frac{T_{CO_2}^2}{T_{CO_2}^1}\right)^n$$

N is less than 1, usually taken about .6 or.7 for petrochemical plants[29].

This is an oversimplification because process equipment and balance of plant components will scale at different rates as shown in Sievert et al, but their variations are not significant for the level of analysis performed here. Using the above cost **scaling** one can get a factor of 4 decrease in costs to match the learning by doing reduction with 5 doublings. Thus, both scaling out for the components of GT technology and scaling up of the processing equipment and balance of plant costs will reduce costs by 4 early on in its deployment. This applies to all DAC technologies with the clear constraint of being on a learning curve path that has a learning curve limit lower than the cost reached. To go from the existing capacity of DAC of kilo tons to gigatonnes is a million fold increase which requires 21 doublings of existing capacity. Thus the above analysis also makes very clear that the low cost will be reached early in the deployment of DAC needed for climate change protection. With the new 45Q subsidies, GT DAC can be economically viable while scaling but as will be made clear below achieving 21 doublings in 20 -30 years is a daunting task that requires mobilization now to be achieved.

The above analysis also makes it clear that from a climate change perspective, the current costs and scale of DAC technologies are relatively less important than their learning cost limit. To date, and because of IP concerns, each DAC effort is pursuing a distinctive path. It also means that one can begin to scale DAC technologies now if they are on the lowest cost path or close to it and there are currently no identified "showstoppers." The remaining challenges for GT are bringing all the high performance that it has demonstrated into a commercially operating large scale unit, solving remaining issues of sorbent lifetime, mechanical reliability, etc. But from a climate perspective, where time is critical one needs to proceed in parallel and not the usual technical readiness step by step process. Ideally of course would be to have a cooperative and coordinated R&D effort on the critical performance parameters to be on a learning by doing path that can achieve low cost.

## Energy Use

The energy demand of GT CDAC is expected based upon the above analysis to be less than 5 gigajoules per tonne of "net" $CO_2$ removed( assuming using renewable energy sources) from the atmosphere for mature technology at scale, including electricity of about 1 gigajoule per tonne. With a need by 2050 of at least 10 gigatonnes of installed DAC CDR capacity to minimize the risk of catastrophic climate change, the new electrical energy required is $10^{19}$ joules or 27,80 TWh. Current world wide electrical generation is of order 29,000 TWh, and this is expected to grow by 2050 to 45,000 TWh , which means DAC electrical consumption would be about 6% of the total. This is a substantial fraction and will require additional investment into energy production from low carbon sources. The use of low carbon energy is critical to the success of DAC because the net $CO_2$ captured per tonne of $CO_2$ directly captured is reduced by the carbon footprint of the energy and this causes the cost and energy per net tonne captured to rapidly escalate. There is an inherent tradeoff between energy use and capital expenditure in DAC which could lead to lower energy use and higher capital systems being deployed, for example by reducing the velocity of the air through the contactor and using even lower quality energy for regeneration.

The low temperature heat needed for GT's CDAC at a scale of 6 doublings will be about 4 gigajoules per tonne. It should be available because electricity production is only 30% of the energy produced with fossil fuels and solar thermal energy sources[30]. Direct production of the heat and electricity needed can be provided by cogenerating solar thermal plants and nuclear sources with DAC $CO_2$ and with the exothermic reactions in using $CO_2$ to make products.

## Conclusion

In conclusion, as a CDR technology, DAC can provide carbon removal at the scale, cost, and time needed to address the climate threat. GT technology is on the lowest cost path identified in the National Academy Study (see Appendix 1), and the performance benefits have been demonstrated by GT though no commercial unit is operating. However, no "showstoppers" have been identified in the large-scale DAC pilot demonstration plants, as all technology's issues, like heat management, sorbent lifetime, and mechanical reliability, need to be addressed. In learning by doing cost reductions, it is also clear, based on the past, that learning includes progress in increasing heat management, sorbent lifetime, and mechanical reliability via continuing R&D . The learning process will result in a low economically viable cost well before the scale needed for climate change.   GT has demonstrated the properties of the $25 per tonne approach identified in the NAS study.

Together with legislation like 45Q, which now provides $130 per tonne of $CO_2$ sequestered in products, the economics of low-cost DAC are very attractive, even while they are at high costs relative to the learning curve limit, which has been shown will be achieved relatively early  in DAC deployment at the scale of millions of tonnes rather than 100's or 1000's of millions of tonnes, thus supporting the Lackner et al conclusion that reaching low cost can be done for a relatively very low cost.

The relatively quick cost scaling will enable DAC to profitably address the climate  challenge we face.

What we need to do is to decide to scale DAC starting now. The technologies needed for it have already been demonstrated but are in need for further R&D while scaling.

 The risk with an increased global effort of not achieving low cost DAC is low and certainly much less than the risk of catastrophic climate change.  The crisis we face is serious enough that if other DAC approaches can demonstrate with data like has been provided above then they too should have a mobilization effort. We believe that other possible low DAC approaches will also benefit from R&D on the critical performance parameters described above. We want to be very clear that we welcome other R&D approaches that show enhanced performance on the critical performance parameters because the focus needs to be on addressing the climate threat.

## Appendix 1

**GT Technology Comparison by the National Academy of Sciences**

Report: [Negative Emissions Technologies and Reliable Sequestration a Research Agenda](#) [15]

The lowest cost process identified by the National Academy directly parallels GT's own patented approach. In other words, without using GT's name (or anyone's) the National Academy of Sciences has said that the approach, technology, and process GT has developed and patented is the best possible approach based on fundamental scientific principles that were described above, to achieve the lowest possible cost for direct air capture. Thus, GT technology is on the learning curve that has the lowest cost limit and by inference all other approaches have limits at substantially higher costs as will be shown below. Notably as described previously, GT has demonstrated the performance that enables the potential benefits of its process to be realized. The learning curve limit of the NAS study of DAC was under $25 and even as low as $18 per tonne.

Throughout the below, relevant portions of the NAS Report that are referenced will be accompanied by the original report page number and placed in quotes. Key portions of these quotes that are particularly germane to the current analysis will be further highlighted in **blue**. Finally, our analysis that compares GT's process to the NAS assessment will be highlighted in **red**.

## Comparison Analysis

**From page 132:** "As has been pointed out by several studies, altering the flow configuration to reduce pressure drop can dramatically reduce capture costs compared to the APS report benchmark system, which is based upon a more conventional approach mimicking post-combustion capture absorber technology. Holmes and Keith introduced a cross-flow scheme for the gas relative to the falling liquid combined with a novel scheme involving the co-capture of $CO_2$ from air combined with an oxy-fired natural gas regeneration in a carbonate-based capture system, leading to reported costs between 336–389 $/t $CO_2$ (Holmes and Keith, 2012b) and 93-220 $/t $CO_2$ (Keith et al., 2018). For solid adsorbents, low pressure drop configurations as preferred to the "honeycomb" structure of monoliths for automobile catalytic converters and other ultra-low pressure drop configurations are preferred motifs (Realff and Eisenberger,2012). These novel configurations will require further testing and demonstration for the lower price points to be realized."

**Commentary**: GT has already demonstrated the potential to achieve the performance in the report that is needed to capture the lower price points.

**From page 147:** "In general, solid sorbent system designs aim to 1) minimize pressure drop for flow through the air-sorbent contactor; 2) minimize contactor mass while maximizing sorbent mass (thus minimizing the sensible heat energy losses); 3) maximize the $CO_2$ uptake; and 4) advantageously manage the water uptake."

**Commentary**:

1. The honeycomb structure enables low pressure drop laminar flow and thus high flow rates of 5m/sec (three to four times faster than other designs), reducing Annualized CAPEX per tonne by that factor.

2. Is developing novel honeycomb contactor structures and is working with honeycomb providers to implement them in ways that maximize sorbent mass on a contactor with low sensible heat. This reduces energy use and thus OPEX

3. GT honeycomb structure has a proprietary coating that maximize $CO_2$ uptake rates (mass transfer rates K and high surface area per volume SAv) per frontal that avoid the deleterious consequences of concentration polarization and by factors of more than three compared to other published approaches

4. Water molecular uptake is not a first order issue for GT technology as opposed to some other sorbent systems because their sorbent does not bind to the H20 molecule. sorbent. Bulk water however is an issue that GT is addressing because it can reduce the kinetics in the pores.

**From page 340** "The desorption time is also defined based on a transient energy balance calculation using 100°C saturated steam as the heat transfer media, which transfers heat through the heat of condensation. In these calculations, the model envisaged that the steam is directly contacted with the sorbent, providing both a concentration and thermal driving force for desorption.

**Commentary**: This is the novel approach that Global Thermostat is using as described above. It allows:

1. The fastest rate of heating minimizes the regeneration time and thus maximizes the amount of $CO_2$ the system can capture per cycle of absorption and regeneration.

2. Increases the driving force for desorption by reducing the partial pressure of $CO_2$ in the pores enabling the temperature driving force to be reduced, allowing for regeneration at temperatures as low a 70C

All this has been demonstrated by GT – enabling 20 second or lower regeneration times and associated 15-minute adsorption times keeping the design objective of 90% in adsorption mode thus maximizing $CO_2$ capture and reducing annualized CAPEX costs.

**From page 150** "An advantage of many recent solid sorbent based manufactured direct air capture processes is that they do not require high temperature thermal energy. In an ideal scenario, the electrical energy needs should be met with renewable energy, and the thermal energy used should be acquired from low temperature waste heat when such heat sources are suitable and available.

"In all scenarios considered in this chapter, the energy used is sourced exclusively for the direct air capture process, and no assumption of "waste heat use is made."

**Commentary**: The cost of energy for the DAC process is based upon the amount of heat needed times the cost of the heat. The GT process reduces both since as noted above 100C heat is generally less costly. In addition, the GT process was patented as a cogeneration process with renewables where the low temperature heat would be available for very low cost from either the power source or the converting of the $CO_2$ into a product. Thus, GT energy costs can be much less than the report considers because low temperature heat will in many cases not be exclusively provided for it but rather obtained within the cogeneration process. This cogeneration process enables very high overall energy efficiency and can be integrated with renewable energy sources, which was the first patent filed by GT.

The report concluded from its cost" analysis on page 152:

"As noted above, the cost estimates span a wide range ($18-$1000 per tonne). Disregarding the lower bound as perhaps not realistically achievable and the upper bound as prohibitively expensive, the middle range of scenarios is perhaps most instructive. These estimates yielded capture costs of 88–228 $/t $CO_2$ for a generic solid sorbent "Direct air capture system."

**Commentary**: As described above, the lowest cost process the National Academy defined (the one leading to $18 per tonne for a million tonne per year facility) is in fact the GT process. While GT is not featured in the report because they decided for proprietary reasons not to publish their results at that time, they have given talks on the general nature of our process and know that several members of the committee understood the benefits and thus contributed to having those benefits featured in the lowest cost approach. Thus, the GT claim of $50 per tonne is supported by the report. In fact, when low temperature heat is available in the cogeneration process, the lower energy costs will enable under $50 per tonne.